\documentclass[aps,preprint]{revtex4}
\usepackage{graphicx}
\usepackage{bm}
\usepackage{cases}
\usepackage{amssymb}
\usepackage{slashed}
\usepackage{color}
\usepackage{subfigure}

\newcommand*{\affmark}[1][*]{\textsuperscript{#1}}

\begin{document}
\title{A coherent structure transport model for scrape-off layer turbulence}

\author{Zhichen Feng\affmark[1] }\thanks{corresponding author's Email: zhichen.feng@colorado.edu}
\author{James Myra\affmark[2]}
\author{Junyi Cheng\affmark[1]}
\author{Calder Haubrich\affmark[3]}
\author{Yang Chen\affmark[3]}
\author{Xinxing Ma\affmark[4]}
\author{Darin R. Ernst\affmark[5]}
\author{Scott Parker\affmark[1]}

\affiliation{\affmark[1]\small{Renewable and Sustainable Energy Institute, University of Colorado, Boulder, Boulder, CO, 80309, USA}}
\affiliation{\affmark[2]\small{Lodestar Research Corporation, Boulder, Colorado, 80023, USA,}}
\affiliation{\affmark[3]\small{Center for Integrated Plasma Studies, University of Colorado, Boulder, CO, 80309, USA,}}
\affiliation{\affmark[4]\small{General Atomics, San Diego, CA, 92121, USA,}}
\affiliation{\affmark[5]\small{Massachusetts Institute of Technology, Cambridge, Massachusetts 02139, USA}}

\begin{abstract}

Understanding the locality of high-temperature plasma energy deposition on material surfaces in fusion reactors is critical for design. Here, we utilize the Gyrokinetic ElectroMagnetic turbulence including X-points (GEMX) simulation, {using ions as tracer particles}, together with SOLPS-ITER solutions for the background equilibrium electric field
including drifts,
to model the heat flux at the divertor plate and characterize the heat load width using realistic X-point geometry. We use a theory-based blobby transport model called the ``Coherent Structure Transport" (CST) model
to include the effect of plasma transport in the edge scrape-off layer.
The CST model is extremely fast and can be used to quickly analyze any SOLPS-ITER solution. SOLPS-ITER provides the steady state, or equilibrium 
on which we superimpose blobby turbulence characterized by blob size, amplitude and frequency. 
We obtain the $1/B_p$ scaling of the heat load exponential decay
width $\lambda_q$, in agreement with 
the Eich empirical scaling and with the Goldston heuristic 
theory. When including blobby turbulence in combination with the SOLPS-ITER electric field, we find a secondary peak in the heat flux radial profile, outwardly displaced from the strike point radius, with a relative amplitude that increases with the initial blob density. We describe the CST model in detail and provide initial investigations of the scaling of $\lambda_q$ and the secondary heat flux peak with blob size and amplitude.
\end{abstract}

\baselineskip 18pt \textwidth6.5in\textheight9in

\maketitle

\section{introduction}





In a tokamak fusion plasma, the heat load at the divertor plate is an important design challenge because the Scrape-Off Layer (SOL) is very narrow radially, resulting in energy flux levels that can exceed material power handling limits\cite{A_Herrmann_2002}. Other magnetic confinement devices face similar power handling challenges\cite{H_Renner_2002}.  
Recent experiments in DIII-D \cite{PhysRevLett.132.235102}, closely matched by gyrokinetic simulations using the XGC code \cite{Chang_2017,10.1063/5.0008755}, show that electron turbulence in the outer pedestal and scrape-off layer can nearly double the width of the divertor heat flux profile, where the measured integral heat flux profile width is proportional to the measured density fluctuation amplitude for electron modes.
First-principles gyrokinetic simulations are extremely time-consuming due to electron transit motion timescales\cite{PhysRevLett.132.235102,Chang_2017,10.1063/5.0008755,Cheng_2023}. 
Here, we simplify the problem by introducing a theoretical description for the blobby turbulence in the SOL region, while still tracking the full gyrokinetic motion of the particles in external fields composed of the magnetic equilibrium, the stationary electric field from SOLPS-ITER and the blobby turbulence. 

Blobby turbulence is often observed in the SOL region in tokamak experiments\cite{10.1063/1.2355668,10.1063/1.3594609,10.1063/1.5018709,stewartzweben,KRASHENINNIKOV_D’IPPOLITO_MYRA_2008,10.1063/1.1426394}. Blobs are localized perpendicular to the B-field to a few ion gyroradii but extended along the field line. The existence of the blobs in the SOL affects the heat flux at the divertor, as we show in this paper. They are difficult to model in gyrokinetic simulation, but can be well described theoretically\cite{KRASHENINNIKOV_D’IPPOLITO_MYRA_2008}. In this paper, we use an analytic blob model\cite{10.1063/1.1426394,10.1063/1.3594609,10.1063/5.0152389} and track {deuterium ions} trajectories with the GEMX code
to investigate the effect of the blob turbulence on the heat flux profile at the lower outer divertor plate. 
We introduce multiple blobs 
upstream
and assume linear superposition of the blob fields, which is a limitation of the model, but for typical parameters, the packing fraction is low enough that there is little overlap of blobs spatially at any point in time\cite{10.1063/5.0021314}.

The paper is organized as follows: Sec.~\ref{intro}
explains in detail how the heat flux profile is calculated in GEMX. Sec.~\ref{Efield} describes how the stationary electrostatic field is implemented, this includes both the time dependent three-dimensional (3D) blob turbulence field and the two-dimensional SOLPS-ITER stationary E-field.
Sec.~\ref{results}  presents the simulation results showing scaling with both blob size and blob amplitude. Sec.~\ref{conclusion and discussion} provides a summary and concluding remarks.


\section{Calculation of the heat flux profile at the divertor plate with the GEMX code}
\label{intro}

GEMX is a gyrokinetic simulation code for modeling tokamak plasmas including the X-point topology of the magnetic field and the SOL region. It uses a structured grid in cylindrical coordinates $(R, Z, \zeta)$ which allows for very fast particle gather-scatter operations and a fast time to solution. At this time, GEMX is electrostatic and all results we present here do not utilize any self-consistent gyrokinetic field solve.  To calculate the energy flux at the divertor plate and heat load width, we use GEMX to advance particle trajectories with the following guiding-center equations of motion\cite{10.1063/1.863594}
\begin{equation}
    \frac{d\mathbf{X}}{dt}= v_{||}\frac{\mathbf{B}^*}{B_{||}^*}+\frac{\mu\mathbf{B\times\nabla B+q\mathbf{\bar E\times\mathbf{B}}}}{qBB^*},
\end{equation}
\begin{equation}
    \frac{dv_{||}}{dt}=\frac{\mathbf{B}^*\cdot\left(q\mathbf{\bar E}-\mu\nabla B\right)}{mB_{||}^*},
\end{equation}
\begin{equation}
    \mathbf{B}^*=\mathbf{B}+\frac{mv_{||}}{q}\nabla\times\hat{b},
\end{equation}
\begin{equation}
    B_{||}^*=\hat{b}\cdot\mathbf{B^*}=B+\frac{mv_{||}}{q}\hat b\cdot\nabla\times\hat{b},
\end{equation}
where $\mathbf X$, $v_{\|}$, $\mu$, $q$, and $m$ are the particle gyro-center position, parallel velocity, magnetic moment, charge, and mass respectively; $\mathbf B$ is the magmatic field at particle position; $B=|\mathbf {B}|$, $\hat b = \mathbf B/B$; $\bar{\mathbf E}$ is the gyro-averaged electric field for the particle's guiding center. 

\begin{figure}[ht]
    \centering
    \includegraphics[scale=0.4]{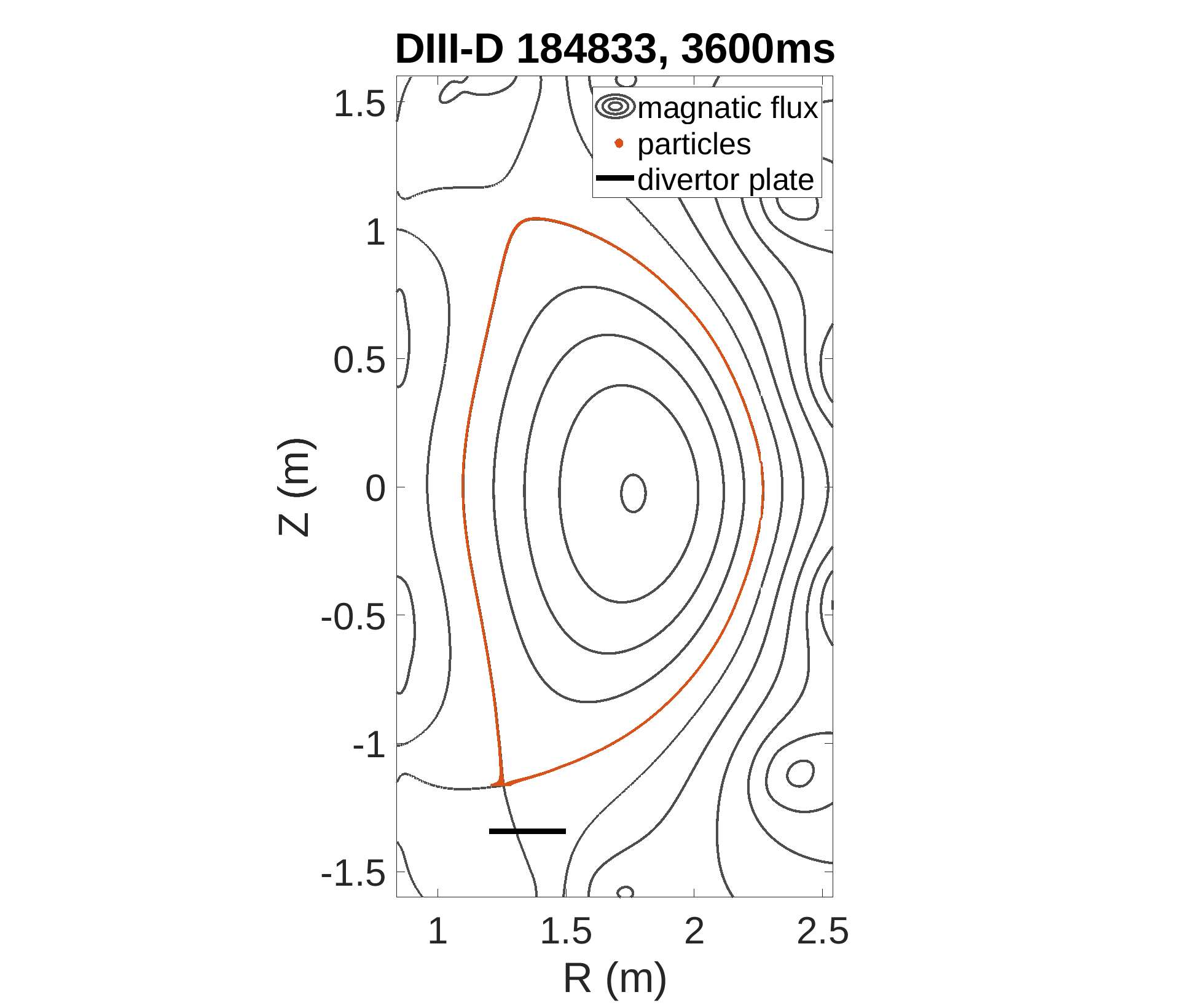}
    \caption{Initial particle load between $(0.999\psi_a, \psi_a)$ in magnetic equilibrium in DIII-D 184833, 3600 ms. Particles appear as red dots which are seen near the separatrix in the figure.  The lower outer divertor plate is also shown.}
    \label{fig_particles}
\end{figure}

To determine the heat flux at the divertor plate, we place particles near the last closed flux surface with uniform {volume} density and temperature distribution, as shown in FIG.~\ref{fig_particles}.
We have verified that the width in $\psi$ of the particle load does not significantly change the results since it is the particles that cross at the last closed flux surface that dominate the energy flux distribution at the plate. {In our model, $\delta\psi_p=0.001$, i.e., the particles are initially deposited in the annulus $(0.999,1)\psi_a$, which is sufficiently small that the particles can be regarded distributed on the last closed flux surface. It also avoids the singularity at the X-points when using a line density. In doing this, we do not model transport occurring inside the separatrix which is not described well using CST.}
Results in this paper employ the magnetic geometry and plasma parameters of the Wide Pedestal Quasi-Coherent H-Mode (WPQH Mode) DIII-D shot 184833 at 3600ms with $T_i=200$eV {for the ion temperature on the last closed flux surface.} \cite{PhysRevLett.132.235102,Ma_2025}. Our goal is to describe the details and demonstrate the utility of the CST model using a realistic and experimentally relevant test case.

If a particle hits the divertor plate (the plate is shown in FIG.~\ref{fig_particles}), it will no longer be advanced and its kinetic energy will be accumulated to the energy flux distribution at that location in $\tilde{s}-\tilde{s}_0$, where 
\begin{equation}
\tilde{s}-\tilde{s}_0=\delta\psi_p/\left|\nabla\psi_p\right|
\end{equation}
is the displacement in minor radius direction (perpendicular to the flux surface) between the position of the particle hitting the divertor and the separatrix, where $\delta\psi_p=\psi_p-\psi_a$ is the difference between the poloidal magnetic flux of the particle hitting the divertor plate and the flux at the separatrix, and $\nabla\psi_{p}$ is the gradient of poloidal magnetic flux at a location to be accumulated. To make a better connection with the heat load width $\lambda_q$, in the literature\cite{Eich_2013}, where the heat flux profile is mapped to the outer mid-plane, we also measure the distance from the separatrix at outer mid-plane by ${s}-{s}_0=\delta\psi_p/\left|\nabla\psi_{p,mid}\right|$, where $\nabla\psi_{p,mid}$ is the gradient of poloidal magnetic flux at the outer mid-plane on the last closed flux surface. 


We utilize the statistics of particles hitting the divertor plate, i.e. a histogram, to calculate the shape of heat flux distribution given by
\begin{equation}
q(s) \propto \left \{ \sum_j v_j^2 \right \}_{s, s+\Delta s},
\end{equation} where $q(s)$ is the heat flux as a function of $s$ and the sum over particles $j$ hitting the plate between $s$ and $s+\Delta s$, and $\Delta s= (s_{\text{max}} - s_{\text{min}})/N_{\text{bins}}$, where $(s_{\text{max}}-s_{\text{min}})$ is the range and $N_{\text{bins}}$ is the number of bins in the histogram. $s_{\text{min}}$ and $s_{\text{max}}$ are determined by where particles hit the plate and $N_{\text{bins}}$ is 100. $N_{\text{bins}}$ along with the total number of particles $N=10^5$, are chosen to obtain good statistics. 
We calculate the mean value of $s$ using
\begin{equation}
\lambda_{avg} \equiv \left< ({s}-{s}_0)\right>= \frac{\displaystyle\sum_i ({s_i}-{s}_0) v_i^2}{\displaystyle\sum_i v_i^2},
\end{equation}
where the sum is now over all particles $i$, that hit the plate. 

Here, $\lambda_{avg}$ is the mean value of $s-s_0$ integrated over the heat flux distribution
\begin{equation}\label{lambda_q_integ}
    \lambda_{avg}=\frac{\int_{-\infty}^\infty (s-s_0) q(s)\text{d}s}{\int_{-\infty}^\infty q(s)\text{d}s}.
\end{equation}
 By integrating the Eich expression for the heat flux\cite{PhysRevLett.107.215001, 10.1063/1.4710517,Eich_2013}, we find the surprising result that that $\lambda_{avg}=\lambda_q$. 
 Namely, the $\lambda_q$ defined in the Eich fit is exactly the heat-flux weighted mean value of $s-s_0$. Details of this result are given in the Appendix \ref{app}.

\begin{figure}[ht]
    \subfigure[The orbits of {ions} hitting the lower outer divertor plate. Different colors indicate different trajectories.]{\label{fig_orbits}\includegraphics[scale=0.25]{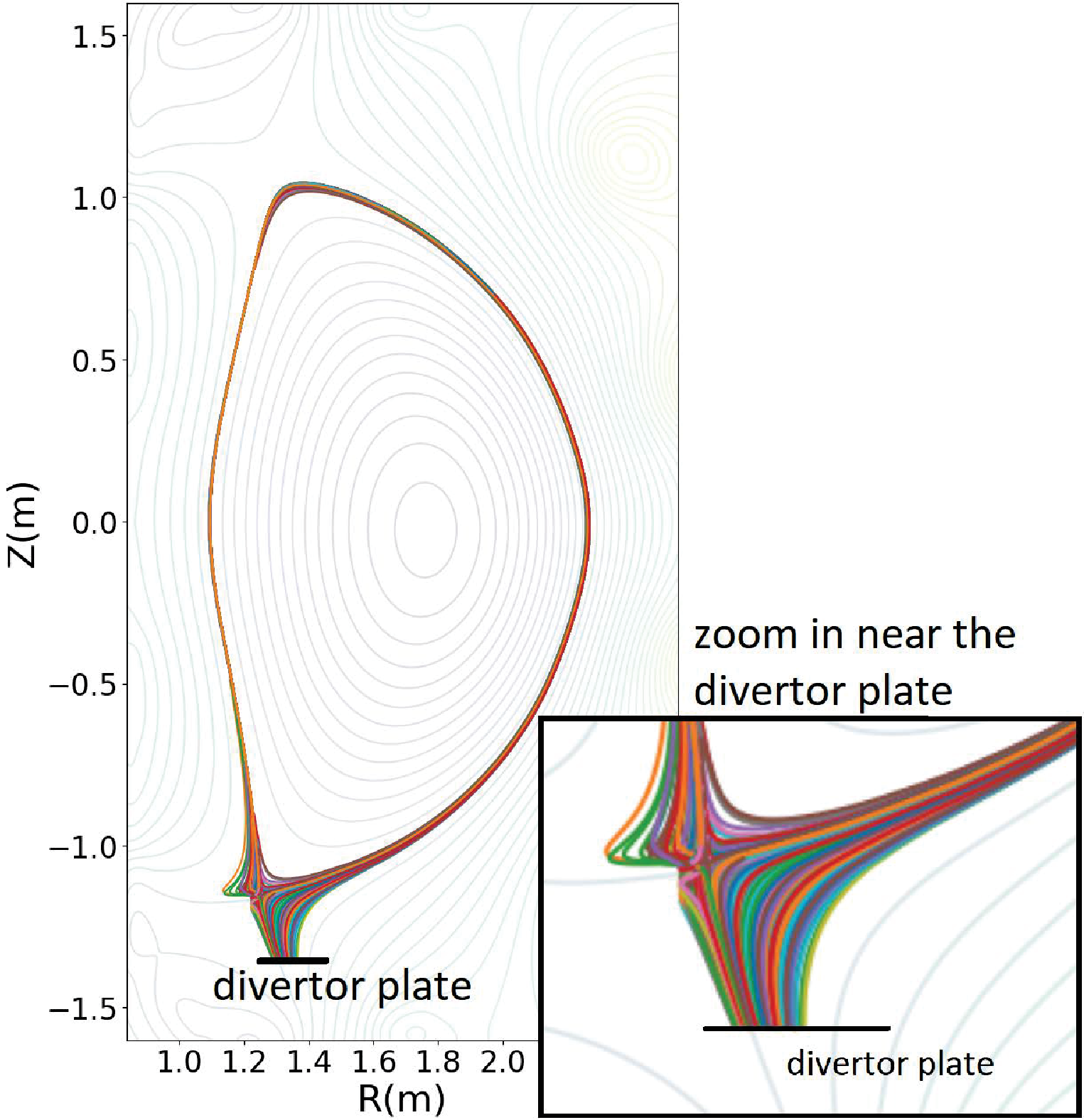 }} 
    \subfigure[{Ion} heat flux on the lower outer divertor accumulated at mid-plane. $s_0$ is the major radius for the last closed flux surface at outer mid-plane.]
    {\label{heat_B0} \includegraphics[scale=0.3]{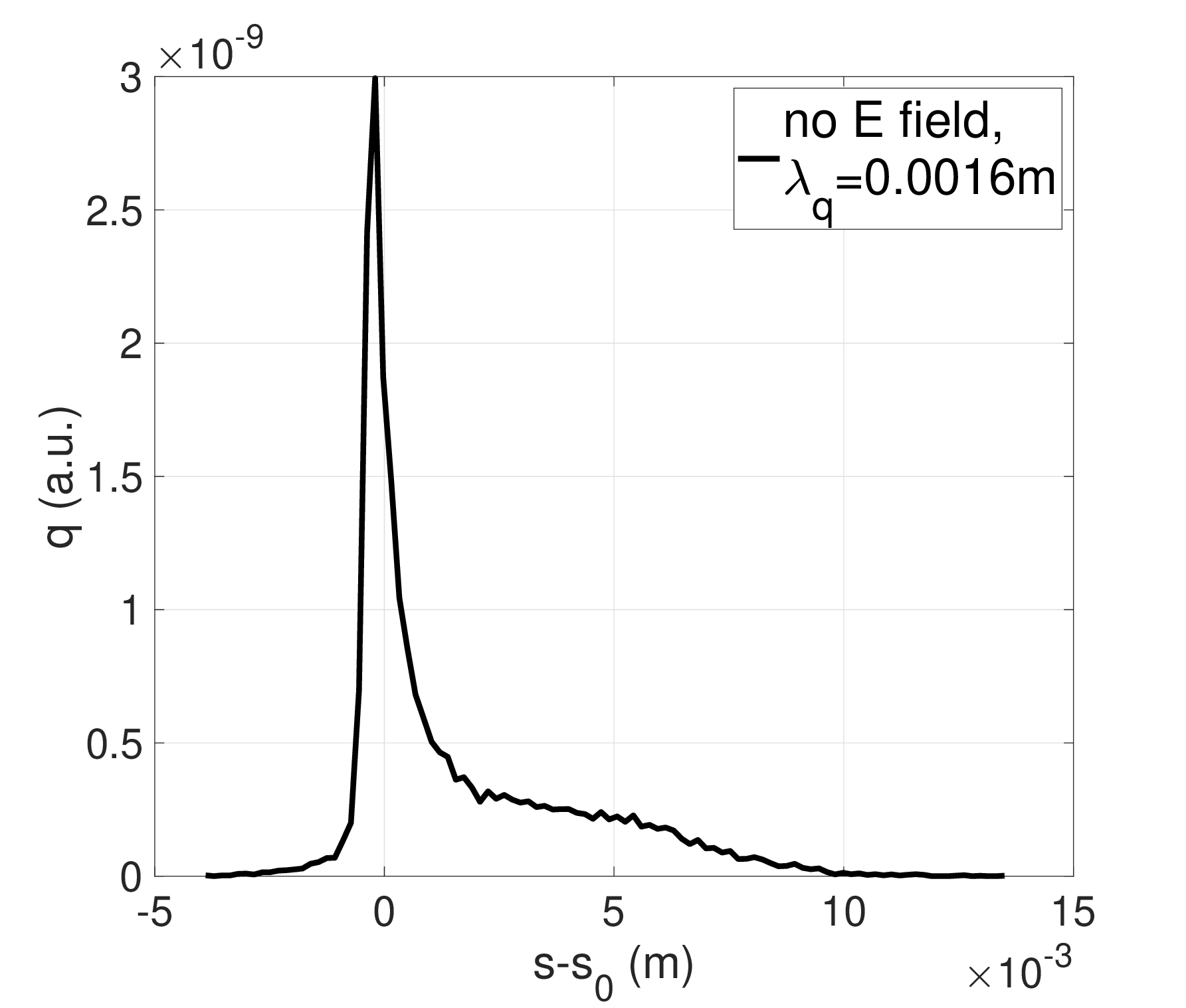}}

\caption{The ion (deuterium) trajectory and heat flux on the lower outer divertor accumulated at mid-plane with the equilibrium magnetic field only.}\label{fig_eq_q}
\end{figure}

FIG.~\ref{fig_orbits} shows the trajectories of deuterium ions  that hit the lower-outer divertor plate. By accumulating the kinetic energy hitting the plate, we can get a heat flux on the lower outer divertor accounted at
mid-plane, as shown in FIG.~\ref{heat_B0}. $\lambda_q$ here is the average distance from the separatrix of the particles hitting the divertor plate mapped back to the mid-plane. The results in FIG.~\ref{fig_orbits} include the equilibrium magnetic field only, and do {\it not} include the electric field from the SOLPS-ITER or the the 3D blobby CST field. The combined effects will be discussed in Sec.~IV.

\section{The E-field in GEMX utilizing SOLPS-ITER and the CST Model}
\label{Efield}

\subsection{The two-dimensional axisymmetric SOLPS-ITER E-Field}
\begin{figure}[ht]
\subfigure[The original electrostatic potential from SOLPS-ITER solution]{
\label{solps00}
\includegraphics[scale=0.2215]{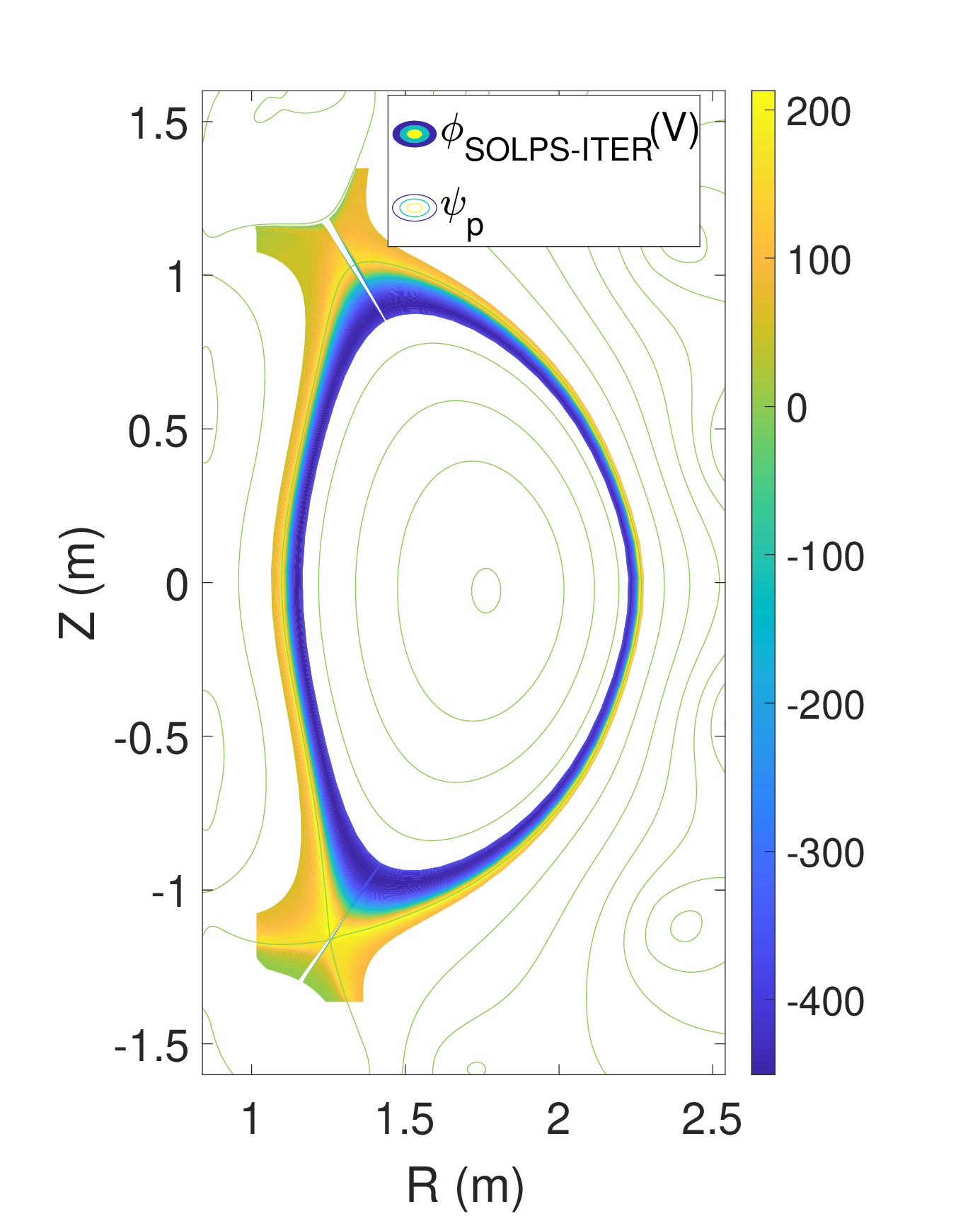}}
\subfigure[The SOLPS-ITER $\phi$ on GEMX structured $(R, Z)$ grid]{
\label{solps_GEMX}
\includegraphics[scale=0.2]{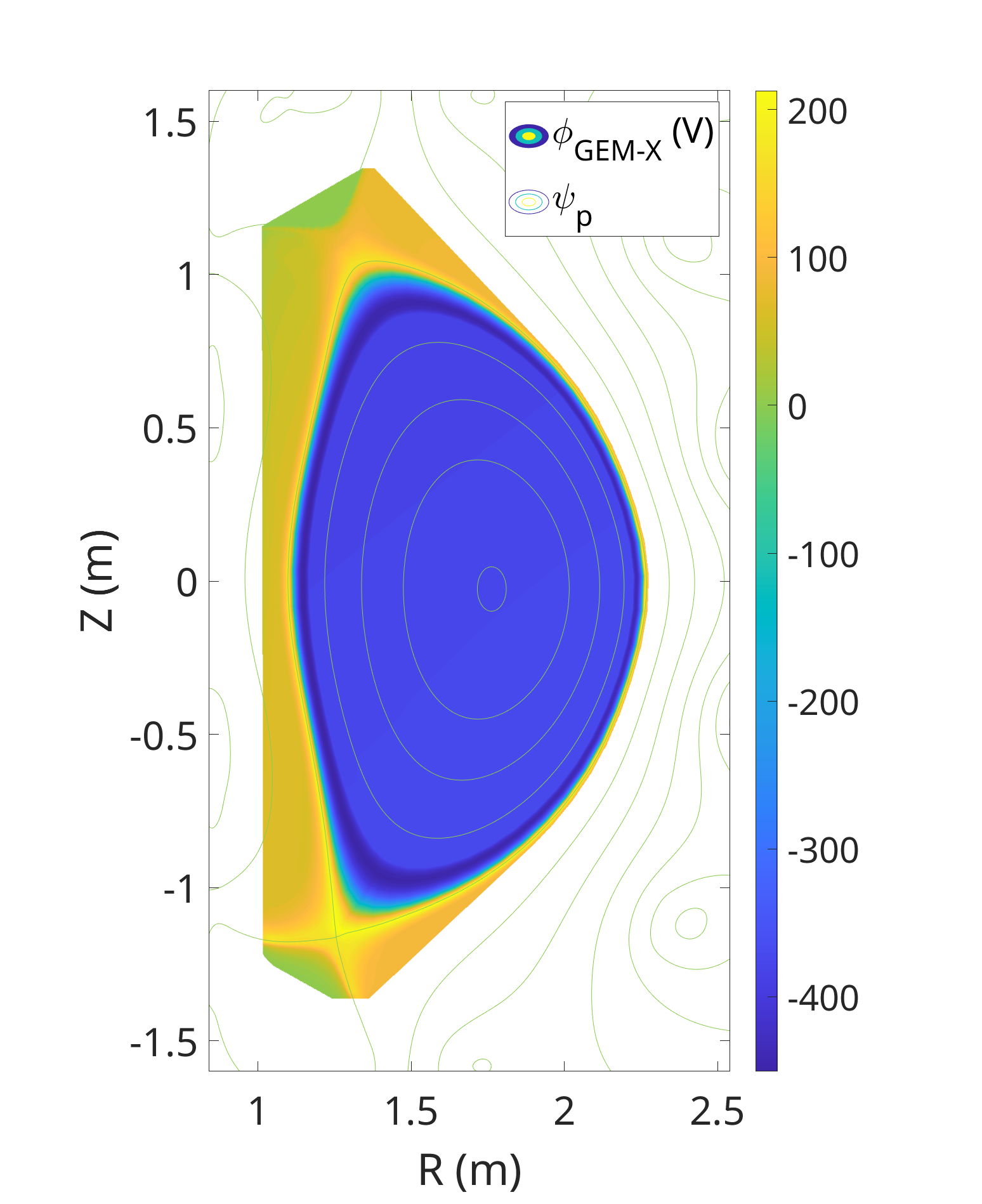}}
\caption{Contour plot of the electrostatic potential from the SOLPS-ITER solution (a), and mapped to the GEMX structured $(R, Z)$ grid (b). }\label{solps_phi}
\end{figure}
In the CST model, we utilize SOLPS-ITER\cite{WIESEN2015480} solutions for the stationary electrostatic potential in the SOL region. SOLPS-ITER is used widely for modeling the SOL/Divertor regions and includes a fair amount of realism including Monte-Carlo neutral modeling. 
The SOLPS-ITER solution we use is the  WPQH Mode DIII-D shot 184833 at 3600ms with $T_i=200$ eV\cite{PhysRevLett.132.235102,Ma_2025} as mentioned previously. The electrostatic potential from SOLPS-ITER is shown in Fig.~3(a), along with the same potential mapped onto the GEMX structured $(R, Z)$ grid shown in Fig.~3(b).

We learned that mapping the block-structured SOLPS-ITER data directly to the GEMX $(R,Z)$ grid leads to spurious $\mathbf{E}\times \mathbf{B}$ orbits and a lack of energy conservation. We determined there was not a sufficient number of grid points along the magnetic flux surface from the SOLPS-ITER solution data to produce a smooth electrostatic potential in $(R,Z)$ using simple spline fitting.  There are slight errors in the numerical calculation of both ${\bf E}_\perp$ and $E_\parallel$. To improve upon this, we linearly interpolate the original data along the flux surface to get more points, then interpolate the data in $(\psi_p,~\theta)$ space onto our structured grid in $(R,Z)$  (FIG.~\ref{solps_GEMX}). The result improves the behavior of the $\mathbf{E}\times \mathbf{B}$ motion and energy conservation.

\begin{figure}[ht]
\subfigure[SOLPS-ITER 2D electrostatic potential at Z=0.]{
\label{SOLPS_Z0}
\includegraphics[scale=0.25]{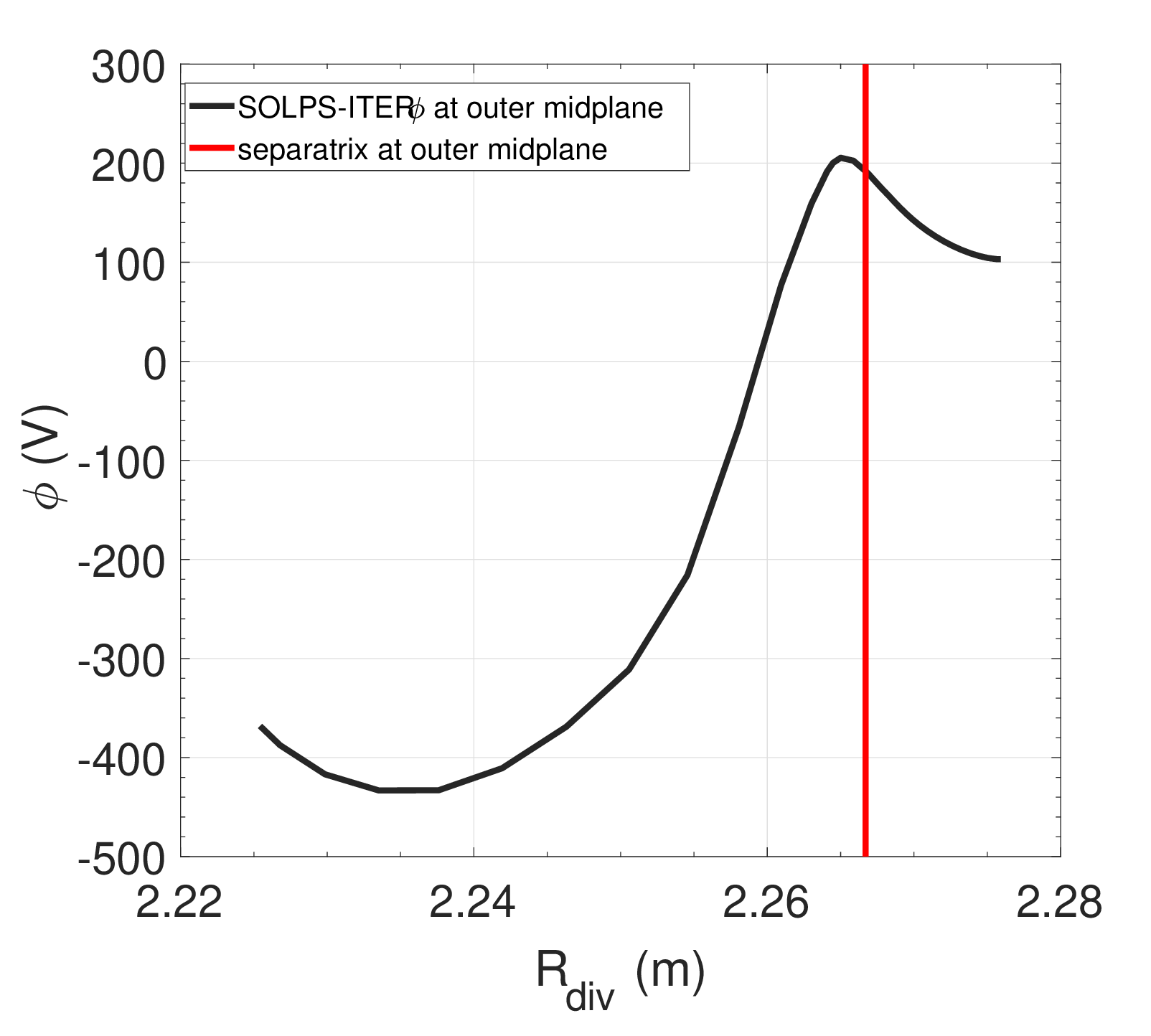}}
\subfigure[SOLPS-ITER 2D electrostatic potential at the lower outer divertor plate.]{
\label{solps_div}
\includegraphics[scale=0.25]{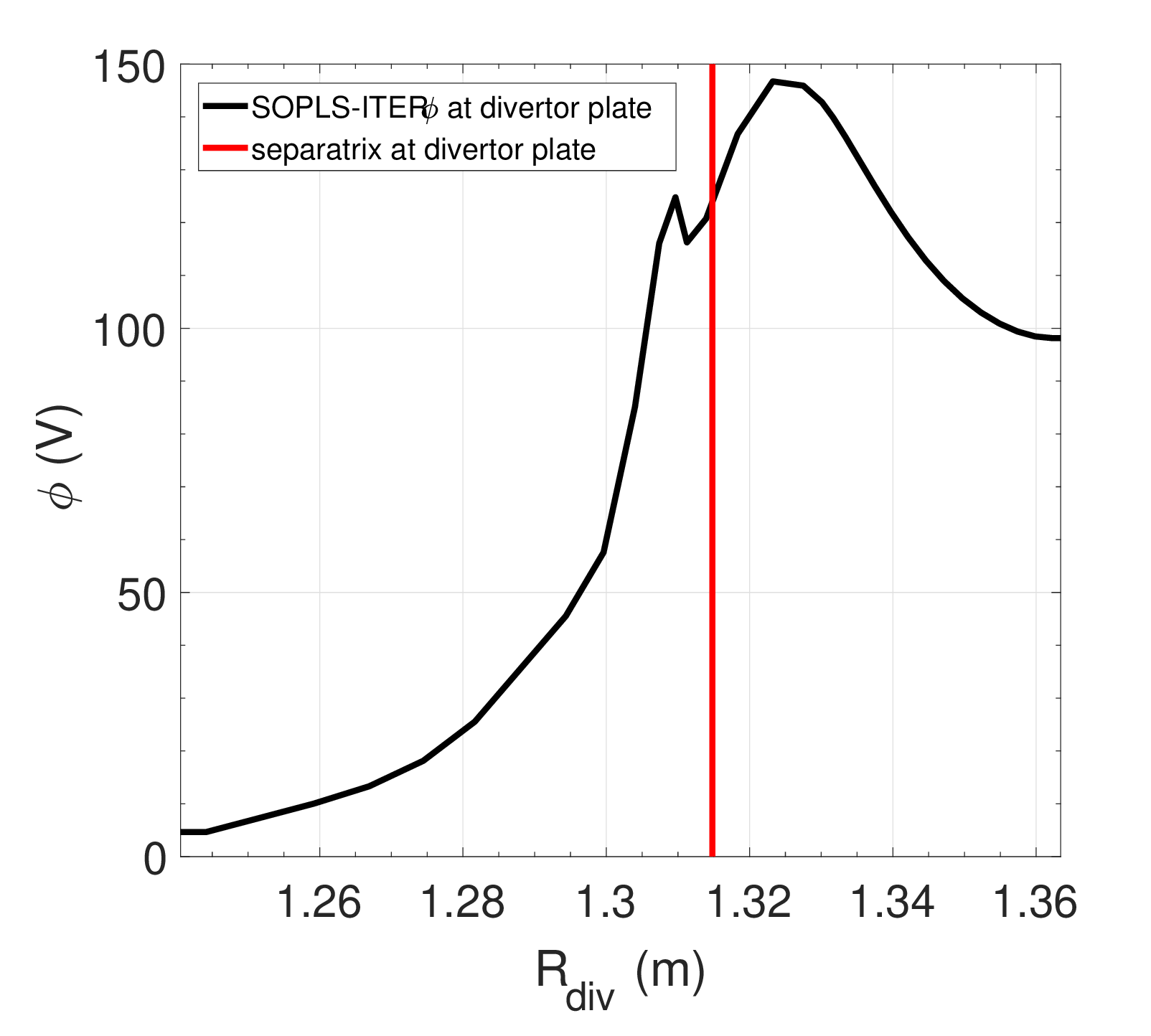}}
\caption{SOLPS-ITER solution of the 2D electrostatic potential near the separatrix both at the outer midplane (a), and at the lower outer divertor (b).}\label{solps_R}
\end{figure}

The SOLPS-ITER field as a function of the major radius at the outer mid-plane and lower outer divertor are shown in FIGS.~\ref{SOLPS_Z0} and \ref{solps_div}. Near the separatrix, the E-field is outward, and particles near this region will drift clockwise due to the $\mathbf{E}\times \mathbf{B}$ drift, and this drift velocity is comparable to the parallel velocity projected on the R-Z plane. This modifies the particle's neoclassical orbit significantly and will make the heat flux profile at the lower outer divertor plate deviate {different} from the Goldston model\cite{Goldston_2012}. There is also a peak in the electrostatic potential outside the separatrix near the divertor plate. {This
peak reduces the energy of the ions approaching the plate at this location.} In the simulation results in Sec.~IV that follow, we will see a dip in the heat flux in this region.

\subsection{The E-field from coherent structures or blobs}

Blobby turbulence is commonly observed in the SOL of tokamak plasmas\cite{10.1063/1.2355668,10.1063/1.3594609,10.1063/1.5018709,stewartzweben,KRASHENINNIKOV_D’IPPOLITO_MYRA_2008,10.1063/1.1426394}. 
The coherent structures are localized perpendicular to B and also have very small $k_{||}$, or are extended along the magnetic field line. We model the blobs as superimposed density perturbations\cite{10.1063/1.3594609,10.1063/5.0152389} of the form 
\begin{equation}\label{blob_model}
     n_b(R,Z,\zeta)=\sum_j n_{jb0}\exp\left [-\frac{\left (R-R_{jb}(l,t)\right )^2}{2\delta_{jR}^2}-\frac{\left (Z-Z_{jb}(l,t)\right )^2}{2\delta_{jZ}^2}-\frac{\left (l-l_{jb0}(t)\right )^2}{2\Delta_{jl}^2}\right ]. 
\end{equation}
 Each individual blob is Gaussian-shaped with half width $\delta_R$ and $\delta_Z$ in the R, Z coordinates, extending along a field line with half width $\Delta_l$. Here, we define $n_{jb0}$ as the amplitude of the particular blob. {$l$ is the arc length with sign of the field line from the point it crosses the outer mid-plane, the sign denotes the direction along or opposite to the direction of the magnetic field. Usually $l_{jb0}(t)=0$ which is at the outer mid-plane.} $R_{jb}(l,t)$ and $Z_{jb}(l,t)$ are the positions of the center of the blob (the field line) at arc length $l$ in the cylindrical coordinates. In our simplified model, we assume the center of the blob propagates across magnetic field lines via the $\mathbf{E}\times \mathbf{B}$ drift velocity\cite{Cheng_2023,10.1063/5.0152389} measured at the mid-plane of the blob 
\begin{equation}\label{dRb}
    \frac{dR_{jb}(0,t)}{dt}=\frac{1}{B^2}E_{Z}(R_{jb}(0,t),0,\zeta)*B_\zeta(R_{jb}(0,t),0,\zeta),    
\end{equation}
\begin{equation}\label{dZb}
    \frac{dZ_{jb}(0,t)}{dt}=-\frac{1}{B^2}E_{R}(R_{jb}(0,t),0,\zeta)*B_\zeta(R_{jb}(0,t),0,\zeta).   
\end{equation}
For a double null configuration, as is the case for DIII-D shot 184833, the blob is sheath-connected and interchange-like. According to the vorticity equation, the electrostatic potential of the blob connecting the two divertor plates (sheath connected in the double-null configuration) can be obtained by\cite{10.1063/1.3594609,10.1063/5.0152389, Cheng_2023}
\begin{equation}\label{blob_potential}
    \phi_b=\frac{2}{e}T_e^{1/2}(T_e+T_i)^{1/2}\frac{L_{||}}{R}\frac{\rho_{s}}{n_0}\frac{\partial n_b}{\partial Z},
\end{equation}
where $T_e$ and $T_i$ are the temperature of electron and ion in the blob respectively; $L_{||}$ is the arc length of the field line that connects the two X-points; $R$ is the major radius at the blob center; $\rho_{s}=m_ic_s/qB$ is the ion Larmor radius calculated with the sound speed $c_s=\sqrt{{T_e}/m_i}$; and $n_0$
is the background plasma density. To simplify the model, we simply use the ion and electron temperatures and plasma density at the separatrix. {In reality, blobs will rapidly cool as they move away from the
separatrix, particularly the electrons, through parallel conduction.}  There are other types of blobs with different forms of electrostatic potential, such as inertial (or resistive ballooning) blobs in single-null tokamak configurations\cite{Tsui,Scotti_2020}. One can use a different expression for the electrostatic potential given in Eq.~(11), depending on the configuration or blob type.

GEMX uses $(R,Z, \zeta)$ coordinates. Like core gyrokinetic simulations, where local $q$ is not too large, such a grid does not accurately represent a structure localized perpendicular to B and field-aligned without a very fine toroidal grid. A nearly field-aligned grid leads to many difficulties near the X-point, typically resulting in the use of a finite-element unstructured grid. To maintain the very simple and fast gather/scatter operations in GEMX, we keep the structured cylindrical grid, and represent the blobs analytically. 

We trace a field line starting from the center of a blob at the mid-plane, which is $\left(R_{jb}(0,t),Z_{jb}(0,t),\zeta_0(t)\right)$, forward and backward for three toroidal turns, the step $d\zeta$ is set to be ${2\pi}/{1280}$ in the toroidal direction, which is sufficient to resolve the field line. Then $R_{jb}(l,t)$, $Z_{jb}(l,t)$, $l(t)$, and toroidal angle are stored in one-dimensional look-up table arrays. 

Each time step, we calculate the position of the blob center at the mid-plane after it moves according to the $\mathbf E\times \mathbf B$ motion, Eqs.~(\ref{dRb}-\ref{dZb}). According to the blob model, Eqs.~(\ref{blob_model}-\ref{blob_potential}), and with the lookup tables, we can get the potential and the $\mathbf{E}$ field of the blob analytically for any particle located at $(R, Z,\zeta)$ quite accurately and efficiently. At the next time step, the blob center moves $(R_{jb}(0,t+dt),Z_{jb}(0,t+dt),\zeta_0(t+dt))$ at the mid-plane according to Eqs.~(\ref{dRb}-\ref{dZb}), and then we re-trace the field line again starting from the new blob center.

\begin{figure}[ht]
    \centering
    \includegraphics[width=0.4\linewidth]{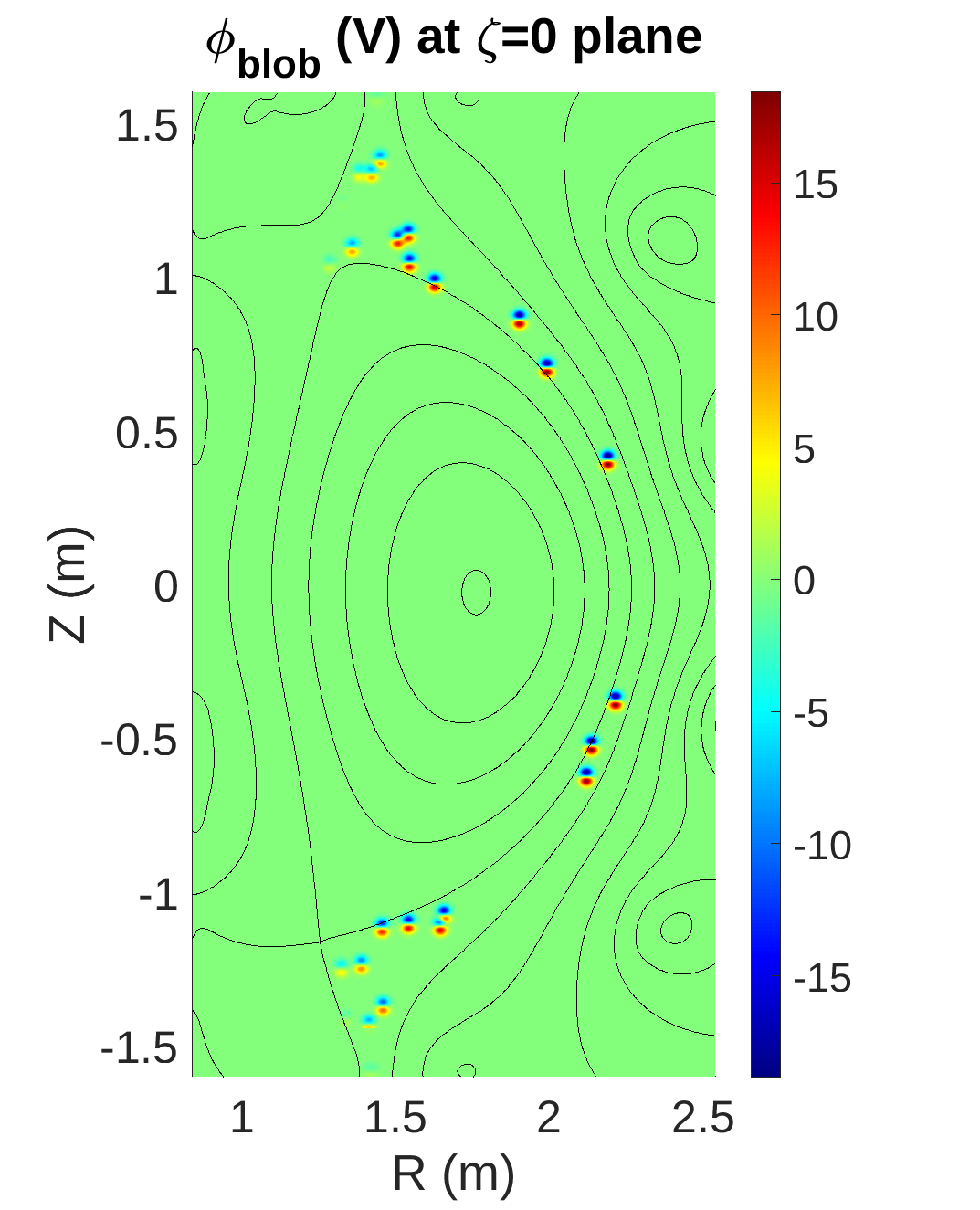}
    \caption{A snapshot of the electrostatic potential of 10 blobs in the equilibrium magnetic field.}
    \label{fig:phi_blob}
\end{figure}

We use typical blob parameters\cite{10.1063/1.5018709,PhysRevLett.108.265001,theodorten,Zweben2022,Militello_2016,Militello2018}. Specifically, we set the amplitude $n_{b0}/n_0 \sim 0.103$, the halfwidth $\delta_{R}\sim\delta_Z\sim 0.015$m and the halfwidth along the field line as $\Delta_l\sim 20$m.  These values are input to the CST model and are typical values observed in experiments. Using these parameters, the electrostatic potential amplitude for each blob is about $e\phi_{b0}/T\sim0.2$ at the mid-plane, where $T$ here is the temperature at the separatrix. The blob will $\mathbf{E}\times \mathbf{B}$ drift outward due to it's own dipole $E_z$ field\cite{KRASHENINNIKOV_D’IPPOLITO_MYRA_2008}. We also use a parameter named $d_{blob}$ to limit how long we track a blob.  If the distance of the blob center from the last closed flux surface reaches $d_{blob}$, the blob vanishes and a new blob is generated at the separatrix. Therefore, $d_{blob}$ is related to the inverse of the blob frequency, assuming a fixed blob velocity. In the results we present here we set $d_{blob}=0.045$ m. The blob packing fraction $f_p$ is the ratio of area covered by the blobs to the total SOL $(R,Z)$ surface area. Since the width of SOL region is much smaller than the blob width, the packing fraction can be approximated by the ratio of the diameter of the blob's half-width to the SOL arc length in the R-Z plane
\begin{equation}
    f_p \approx N_{blob}N_{turns}\frac{2\sqrt2r_{blob}}{L_{SOL}}\frac{\sqrt{2}r_{blob}}{d_{blob}},
\end{equation}
where $N_{blob}$ is the total number of blobs, $N_{turns}$ is the number of times the field-aligned extended blob crosses the toroidal plane, $r_{blob}$ is the width of the blob in R-Z plane, $L_{SOL}$ is the arc length of the last closed flux surface in R-Z plane connecting the two X-points. $f_p=0.0125$ for one blob in the system with typical values of the parameters: $N_{turns}=2$, $r_{blob}=0.015$m, $L_{SOL}=3.2$m, and $d_{blob}=0.045$m. 

FIG.~\ref{fig:phi_blob} shows a snapshot of the electrostatic potential for 10 blobs for the WPQH mode DIII-D shot 184833 at 3600 ms. For illustration, we have increased the blob size to $\delta_{R}\sim\delta_Z\sim 0.02$m, to make the blobs more visible. No SOLPS-ITER equilibrium electrostatic potential is included in FIG.~\ref{fig:phi_blob}. Note that there are more blob crossings of the plane near the X-points due to the nonlocal variation of $q$ (poloidal variation). The poloidal magnetic field is very weak near the X-points and a near flute-like blob structure crosses the R-Z plane more times in these regions.

\section{Divertor heat load calculations using the CST model}
\label{results}


In this section we present results using the CST model in the GEMX code. We will present three cases.  First, $\lambda_q$ scaling using only the magnetic equilibrium will be investigated as a benchmark with the typical $1/B_p$ scaling\cite{Goldston_2012}. Second, we add the SOLPS-ITER axisymmetric E-field (without blobs) to study it's effect on the heat flux distribution. Finally, we introduce the blobs using the CST model and show results with both blobs and the equilibrium E-field coming from SOLPS-ITER. Results for all cases again use WPQH Mode DIII-D shot 184833 at 3600 ms with $T_i=200$ eV\cite{PhysRevLett.132.235102,Ma_2025}. 

\subsection*{Case I: Equilibrium B-field field only}

\begin{figure}[ht]
    \centering
    \includegraphics[scale=0.3]{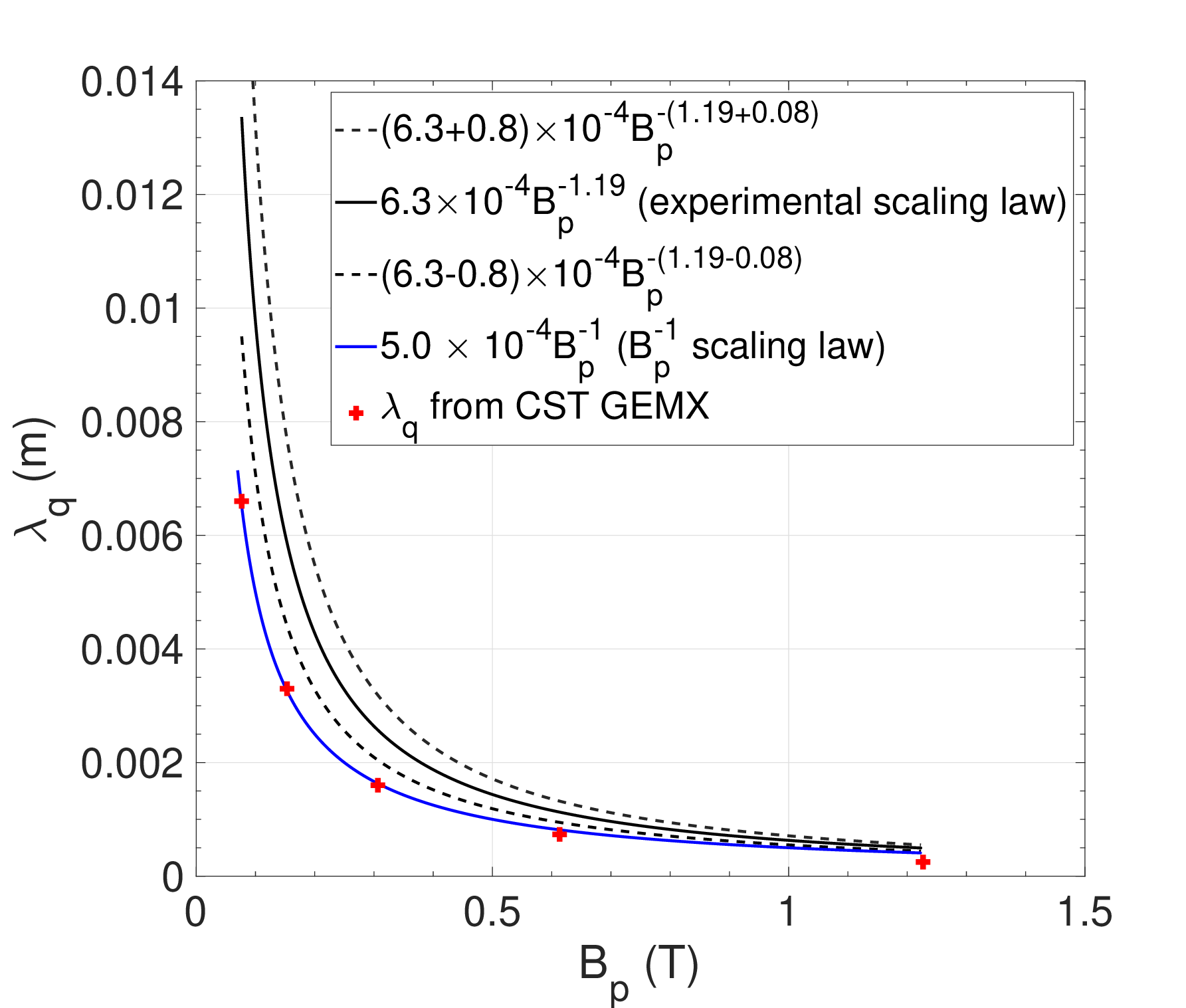}
    \caption{$\lambda_q$ versus the poloidal magnetic field strength $B_p$ at the mid-plane using the equilibrium magnetic field only.}
    \label{fig:lambda_q vs Bp}
\end{figure}

We begin by showing results with the equilibrium magnetic field only. We see that the conventional $1/B_p$ scaling of the heat load width $\lambda_q$ is obtained.  We load deuterium ions with a shifted Maxwellian distribution in $v_{||}$. We add a parallel flow $c_s$ towards the lower outer divertor to account for the plasma pre-sheath acceleration. FIG.~\ref{fig:lambda_q vs Bp} shows the scaling law of $\lambda_q$ versus $B_{p}$ (the poloidal magnetic field at outer mid-plane here). In this figure, we find that $\lambda_q\sim0.5B_{p}^{-1}$(mm) calculated by GEMX. 

The scaling law $1/B_p$ is expected due to the decrease of the drift orbit width.  We note that $\lambda_q$ from GEMX with the equilibrium B field only is below the empirical scaling law from experiment\cite{Eich_2013}, which is $(0.63\pm0.08)B_{p}^{-(1.19\pm0.08)}$(mm). Lack of agreement is
not surprising because neither the axisymmetric equilibrium E-field or turbulence is present.
To obtain a realistic scaling, further work is needed to include the $B_p$ scaling of the SOLPS-ITER axisymmetric E-field (a SOLPS-ITER $B_p$ scan), as well as the scaling of the blob turbulence with $B_p$.

\subsection*{Case II: SOLPS-ITER axisymmetric E-field and the equilibrium B-field}

The equilibrium electric field near the separatrix as calculated by SOLPS-ITER is quite large, as shown in FIG.~\ref{SOLPS_Z0}. The resulting $\mathbf E\times \mathbf B$ velocity in the poloidal direction generated by $E_r$ at mid-plane is large enough to be comparable to the projection of parallel velocity in the poloidal plane. This will change the neoclassical orbit significantly. Hence, the SOLPS-ITER electric field is needed for a realistic model in GEMX.  Calculation of the axisymmetric equilibrium E-field self-consistently in GEMX using conventional particle-in-cell simulation is ongoing work. SOLPS-ITER provides a more accurate description at this time.

\begin{figure}[ht]
    \centering
    \includegraphics[scale=0.35]{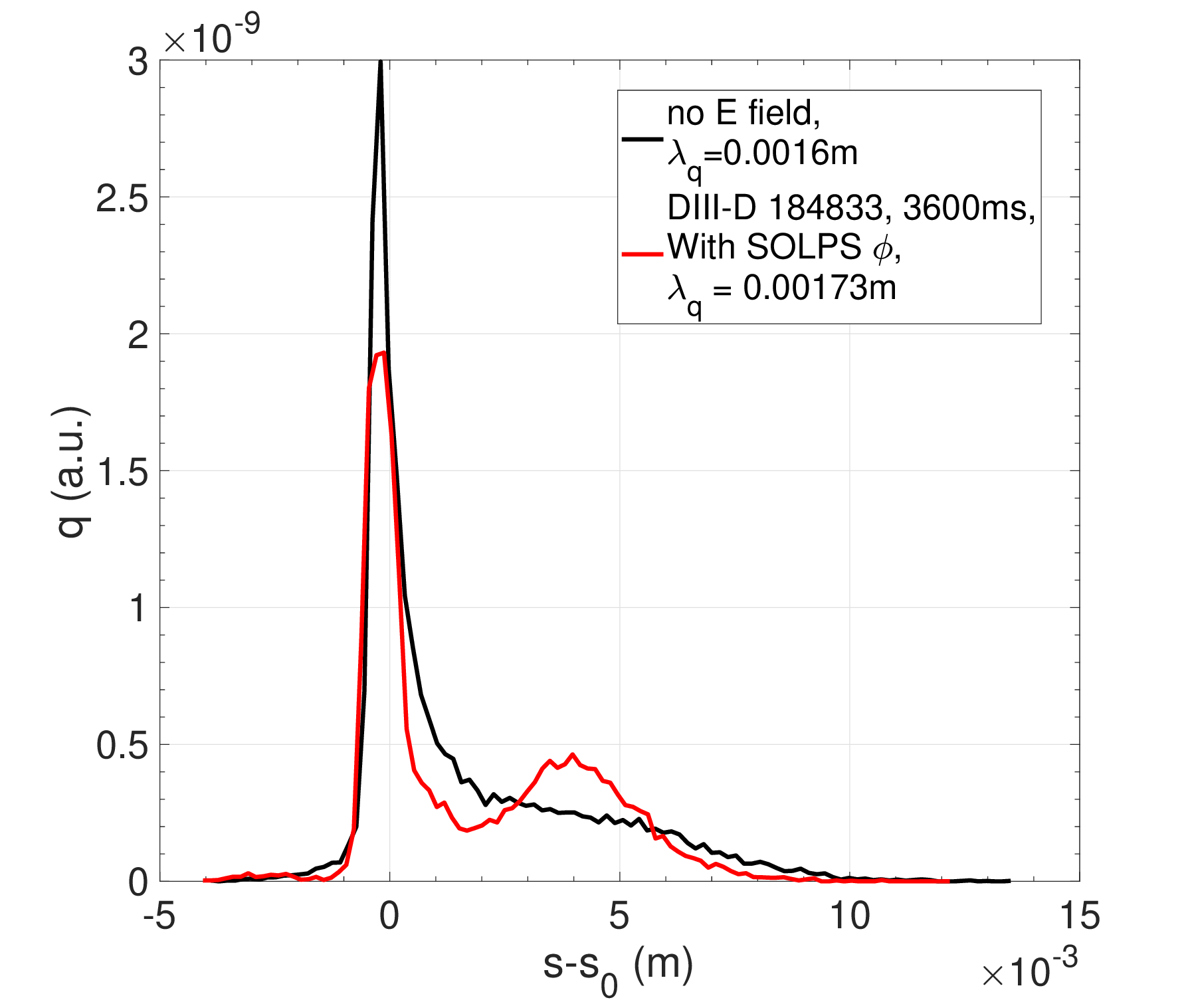}
    \caption{Heat flux profiles mapped back to the outer midplane for Case I (equilibrium B-only) and Case II (SOLPS-ITER E-field and equilibrium B-field).}
    \label{fig:q_eq_SOLPS}
\end{figure}
FIG.~\ref{fig:q_eq_SOLPS} shows the heat flux profile mapped back to the outer midplane including the SOLPS-ITER axisymmetric electric field. The result with the equilibrium B-field only is shown for comparison.  
{There is a secondary peak in the heat flux shown in FIG.~\ref{fig:q_eq_SOLPS}, which is
sometimes observed in experiment\cite{C.J.Lasnier_1998, PhysRevLett.124.195002, 10.1063/5.0048609}. The non-monotonic energy flux density profiles
at the divertor plate were also found in ITER modeling studies\cite{Moulton_2026}.}
{As discussed in Sec.~III, 
there is an electric field peak at the lower outer divertor, see FIG.~\ref{solps_div}, and this  appears to lower the heat flux in this region at the plate.} The heat flux is broadened and forms a second peak due to the axisymmetric equilibrium electric field. 

\subsection*{Case III: Blobby turbulence, SOLPS axisymmetric E-field, and equilibrium B-field}

\begin{figure}[ht]
    \centering
    \includegraphics[width=0.5\linewidth]{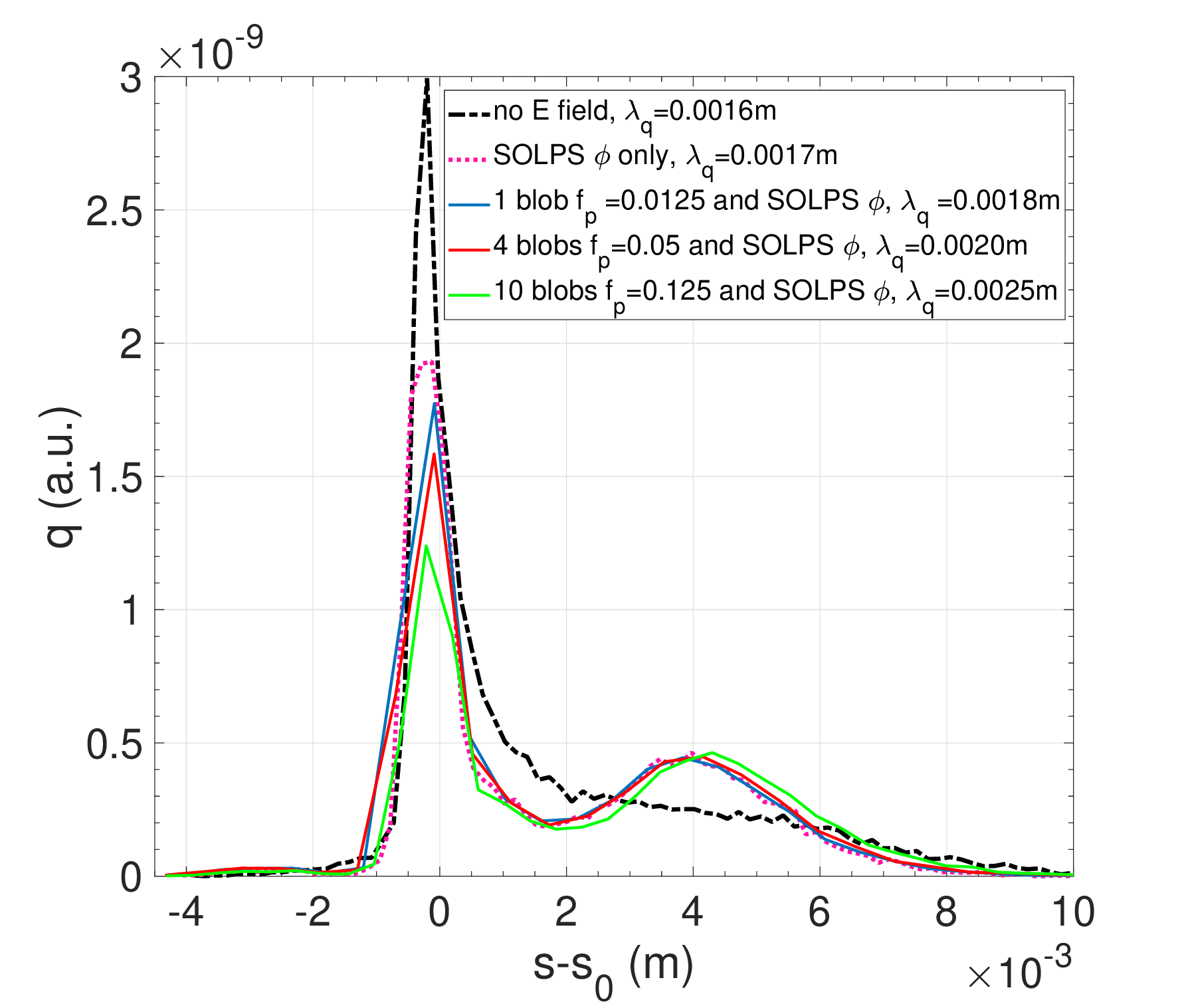}
    \caption{Heat flux profiles at the lower outer divertor plate for five cases with added realism (see legend). For each case the calculated value of $\lambda_q$ is also given. The cases include the equilibrium B-field only, the equilibrium B-field and the SOLPS-ITER E-field, and blob(s) with the SOLPS-ITER E-field and the equilibrium B-field.}
    \label{fig:q_all}
\end{figure}

FIG.~\ref{fig:q_all} shows the heat flux profile at lower outer divertor plate for five specific cases with added realism: the equilibrium B-field only; the equilibrium B-field along with the axi-symmetric SOLPS-ITER E-field without blobs; and cases with 1, 4, and 10 blobs (packing fractions of 0.0125, 0.05, 0.125 respectively) with the SOLPS-ITER E-field and the equilibrium B-field. The blob parameters are $n_{b0}/n_0 = 0.103$, $\delta_{R} = \delta_Z = 0.015$m, $d_{blob}=0.045$m, $\Delta_l=20$m. 
The blobs are distributed uniformly in the toroidal direction. 
We observe that the blob turbulence will broaden the heat flux profile and reduce the heat flux peak at the separatrix. With 10 blobs, the heat flux peak at the strike point will be reduced by about one-half. 

\begin{figure}[ht]
    \centering
    \includegraphics[width=0.5\linewidth]{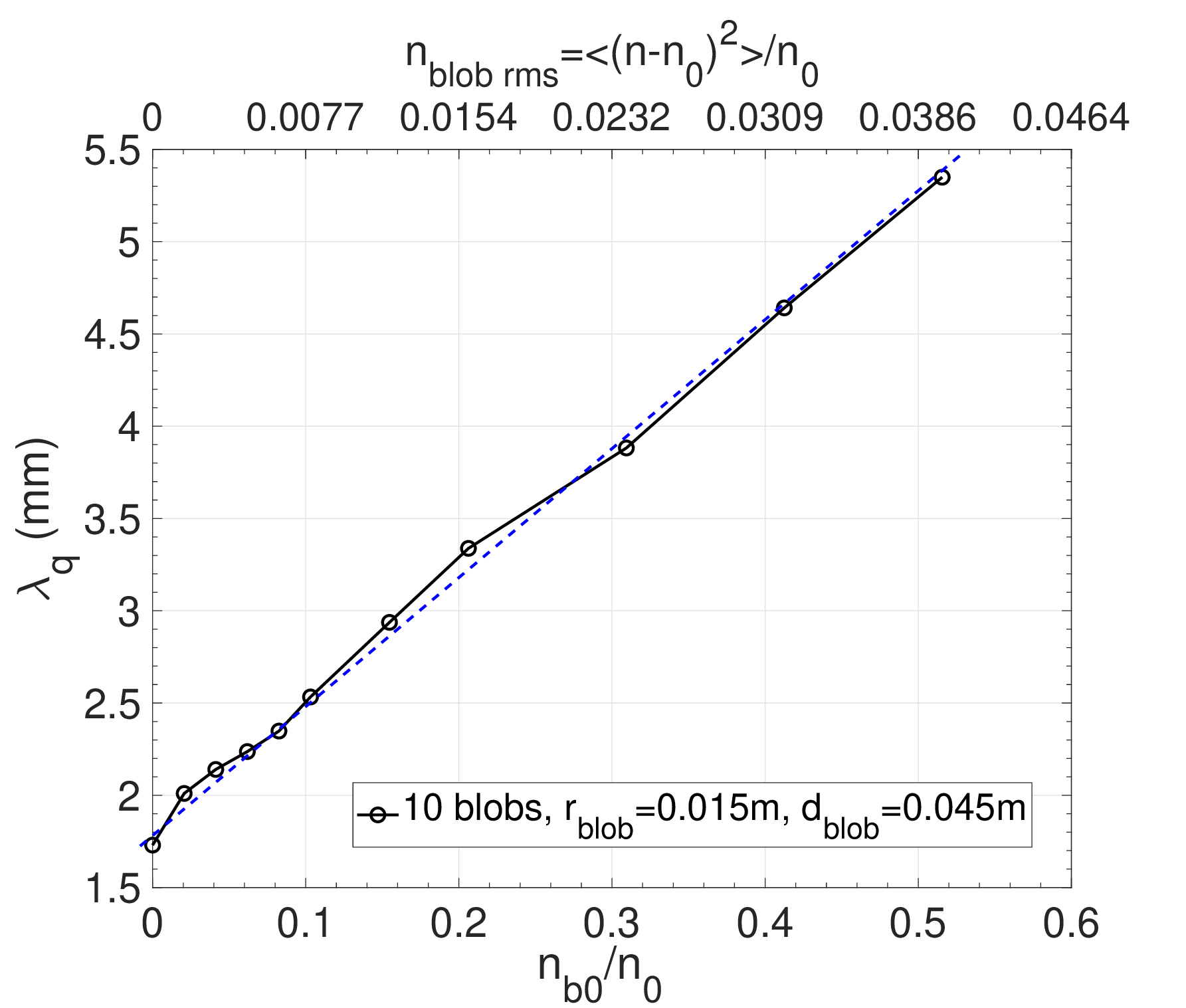}
    \caption{$\lambda_q$ versus blob amplitude. Note that the perturbed density is normalized to the density $n_0$ at the separatrix, which will make the apparent blob perturbation amplitude much smaller than the actual value. The blue dashed line shows linear scaling.}
    \label{fig:lambda_vs_A}
\end{figure}

FIG.~\ref{fig:lambda_vs_A} shows the $\lambda_q$ scaling with blob amplitude using ten blobs. The blob amplitude, the y-axis in
FIG.~\ref{fig:lambda_vs_A}, is expressed two ways. At the bottom of the plot  the blob center density over the background density $n_{b0}/n_0$ is shown. At the top, the normalized root mean square (RMS) of the perturbed density\cite{Zwebenblobrms} is shown, specifically $\delta n_{blob~rms}/n_0=\left< (n-<n>)^2\right> ^{1/2}/n_0$ at the mid-plane. 

Here, we use $n_0$ at the separatrix to normalize $\delta n$ so the blob amplitudes $n_{b0}/n_0$ may seem smaller than $O(1)$.  If normalized to the local background density they would be order unity. 
We find that {additional $\lambda_q$ broadening} is very nearly proportional to the blob amplitude and that blobby turbulence can significantly broaden the heat load width.  For example, for these parameters, $n_{\mbox{blob rms}}/n_0$ of a few percent (as measured at the outer midplane) of blobby turbulence can double $\lambda_q$, which is similar to experimental observations\cite{Eich_2020}.

\begin{figure}[ht]
    \centering
    \includegraphics[width=0.5\linewidth]{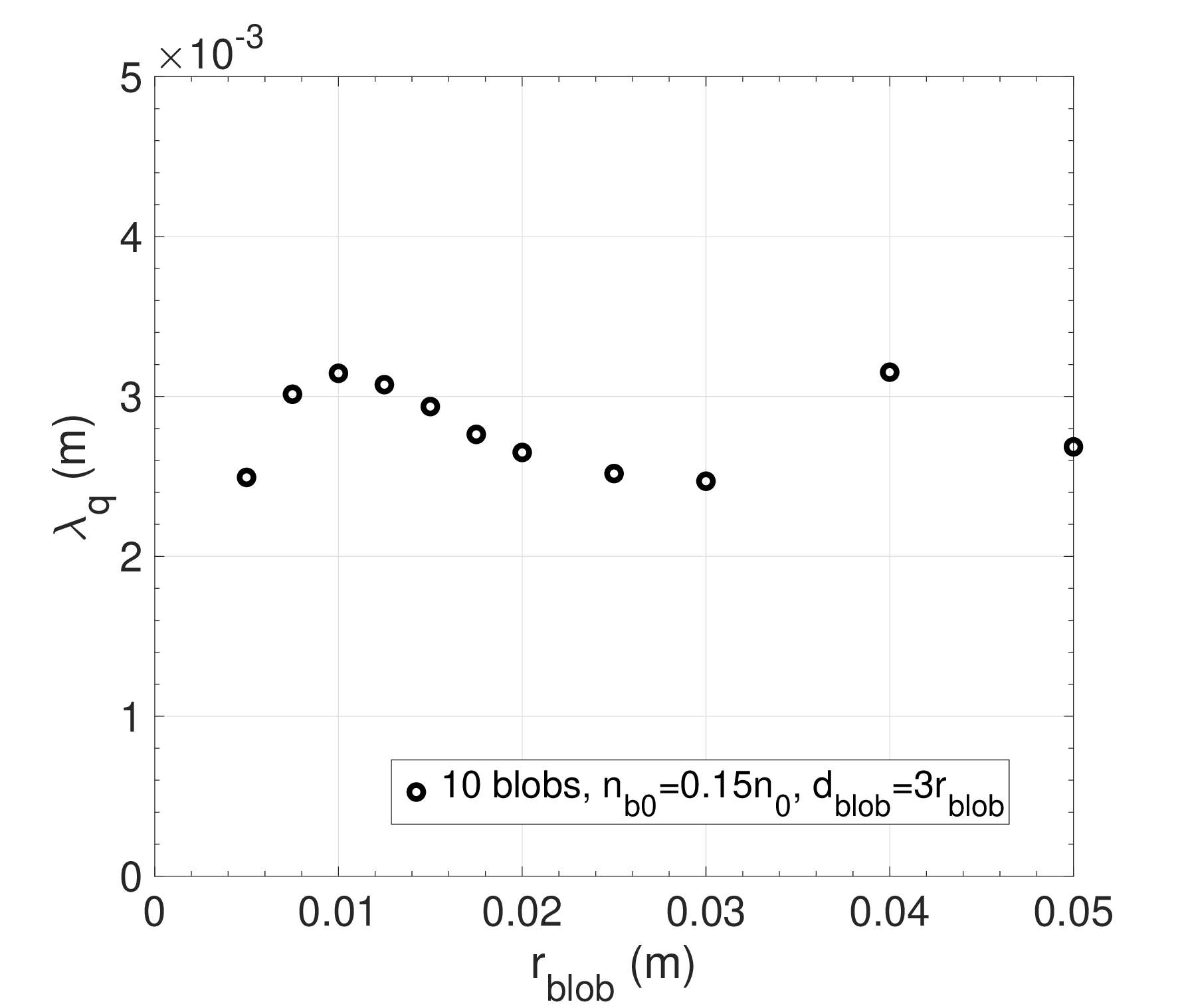}
    \caption{$\lambda_q$ versus the blob size, $r_{blob}$, keeping $d_{blob}=3r_{blob}$ to make the packing fraction proportional to $r_{blob}$.}
    \label{fig:lambda_rb}
\end{figure}

FIG.~\ref{fig:lambda_rb} shows scaling of the heat load width, $\lambda_q$, with blob size. Results are shown for a case with 10 blobs. The blob amplitude is fixed in
FIG.~\ref{fig:lambda_rb} with $\delta n_{b0}/n_0=0.15$. $d_{blob}$ is set to $3r_{blob}$ which keeps the packing fraction roughly constant with varying blob size. We find that $\lambda_q$ only changes modestly ($2.5\sim3.2$mm) 
over an order of magnitude variation of blob size (holding packing fraction constant). This can be explained by the fact there are competing effects.  Increasing the width of blob causes more interaction with the blob.  However, increasing the blob size decreases the blob's electrostatic potential as seen from
Eq.~(\ref{blob_potential}).

The calculated heat flux width using the CST model in GEMX, including the magnetic equilibrium, the SOLPS-ITER field and blobby turbulence is $\lambda_q=2.9$mm for this DIII-D WPQH mode test case. We used typical SOL blob parameters ($r_{blob}\sim0.015$m, packing fraction$\sim 0.125$, and $\delta n_{rms}/n_0\sim0.013$). The calculated $\lambda_q$ from our model is similar to the experimental results\cite{PhysRevLett.132.235102} and XGC1 gyrokinetic simulation results\cite{PhysRevLett.132.235102} for the DIII-D WPQH mode.

\section{Summary and future work}
\label{conclusion and discussion}

The CST model in the GEMX code (GEMX/CST) provides a relatively simple way to obtain heat flux profiles at the divertor plate. The model includes realistic X-point magnetic geometry, an axisymmetric stationary E-field calculated using SOLPS-ITER and a theory-based blobby turbulence model.  GEMX/CST can quickly obtains a physical heat load width. Typically, less than 10 minutes is needed for each run on one node of the Perlmutter supercomputer at the National Energy Research Supercomputing Center (NERSC).
GEMX/CST shows that the stationary axisymmetric SOLPS-ITER E-field has a significant effect on the ion orbit dynamics and the associated heat flux distribution on  the lower outer divertor plate. In fact, the radial electric field is large enough compete with with the parallel flow in the poloidal direction. The effect is to broaden and produce a secondary peak in the {ion} heat flux radial distribution at the plate. Additionally, we find that blobby turbulence can further reduce the {ion}  heat flux level and broaden the heat flux radial distribution. The {ion} heat load width can be approximately doubled by including blobby turbulence with $\delta n_{blob~rms}/n_0 \sim 0.023$.

Results presented in this paper include ions only. Future work will include incorporating electron physics. Currently, GEMX/CST without a sheath boundary condition at the plate, vastly overestimates the electron heat flux. The electron contribution to the heat flux at the divertor plate is reduced by the negative sheath potential near the divertor plate. The sheath potential drop will return electrons within the bulk of the $v_\parallel$ distribution and de-accelerate the electrons that do hit the plate. Future work will include implementation of a logical sheath boundary condition\cite{PARKER199341} for electrons.  The heat flux from electrons will be distributed within a very narrow layer, relative to ions, because of the smaller electron orbit width. Therefore, accounting for the electrons is still important. Understanding how the electrons interact with the blobs will be a valuable use of the GEMX/CST model. 

Here we define the basic GEMX/CST model and show results for the WPQH mode on DIII-D. Much more work is needed to see how $\lambda_q$ scales with parameters using multiple experimental instances (both shot number and time instance). Charge-neutral collisions need to be considered. One complication with GEMX/CST is that to do a realistic parameter scan, a SOLPS-ITER solution is required for each instance, which is not always available.  However, with the availability of the GEMX/CST model, better coordination with existing and ongoing SOLPS-ITER studies may lead to a better understanding of heat load scaling.

\appendix*
\section{Relationship between $\lambda_{avg}$ and $\lambda_q$}\label{app}
The calculation of $\lambda_q$ is facilitated by determining $\lambda_{avg}$ in Eq.~(\ref{lambda_q_integ}).  Here, we show that $\lambda_{avg}=\lambda_q$, where $\lambda_q$ is given by Eich\cite{PhysRevLett.107.215001, 10.1063/1.4710517,Eich_2013}. This is a nontrivial result because $\lambda_{avg}$ is mean distance from the separatrix weighted by the heat flux distribution.
The Eich expression for the heat flux is\cite{Eich_2013}
\begin{equation}\label{A1}
    q(s-s_0)=\frac{q_0}{2}\exp\left(\left(\frac{S}{2\lambda_q} \right)^2-\frac{(s-s_0)}{\lambda_q} \right)\cdot \text{erf}c\left(\frac{S}{2\lambda_q}-\frac{(s-s_0)}{S} \right)+q_{BG},
\end{equation}
and $q_{BG}$ is the background DC heat flux and is zero in the GEMX/CST model. We make the following substitutions: $\mu=S/(2\lambda_q), \sigma= (s-s_0)/S$. Eq.~(\ref{A1}) then
becomes 
\begin{equation}\label{A2}
    q(\sigma)=\frac{q_0}{2}\exp\left(\mu^2-2\sigma \right)\cdot \text{erf}c\left(\mu-\sigma\right).
\end{equation}
We now use Eq.(\ref{lambda_q_integ}) to calculate $\lambda_{avg}$ 
\begin{equation}\label{A3}
    \lambda_{avg}=\frac{\int_{-\infty}^\infty (s-s_0)q(s)\text{d}s}{\int_{-\infty}^\infty q(s)\text{d}s}=\frac{S\int_{-\infty}^\infty \sigma q(\sigma)\text{d}\sigma}{\int_{-\infty}^\infty q(\sigma)\text{d}\sigma}.
\end{equation}
Performing the integrals in Eq.~(\ref{A3}), the numerator is 
$(q_0/2)\times S/(2\mu^2)$
and the denominator is $(q_0/2)/\mu$. Therefore, 
\begin{equation}\label{A4}
    \lambda_{avg}=\frac{S/(2\mu^2)}{1/\mu}=\frac{S}{2\mu}=\lambda_q.
\end{equation}

\begin{acknowledgments}
This research was supported by the Frontiers in Leadership
Gyrokinetic Simulation project, Scientific Discovery through
Advanced Computing program, U.S. Department of Energy (DOE)
Contract DE-SC0024425, and by the U.S. Department of energy Fusion Innovation Research Engine Collaborative, ``Advanced Profile Prediction for Fusion Pilot Plant Design (APP-FPP)", via MIT subcontract under award No. DE-SC0025853.

\end{acknowledgments}

\bibliography{ref}

\end{document}